\documentclass[pra,article]{revtex4}

\newcommand{\rmi}{\operatorname{\rm i}}

\usepackage{amsfonts}
\usepackage{amssymb}
\usepackage{amsmath}
\usepackage{bbm}
\usepackage[]{graphicx}

\begin{document}

\title{\Large On quantum bound states in equiperiodic multi-well potentials I:\\ {\large $-$  Locally periodic potential sequences and Floquet/Bloch bands.}}

\author{Karl-Erik Thylwe $\vspace {0.3cm}$}

\affiliation {Kl\"{o}verv\"{a}gen 16, 387 36 Borgholm, Sweden \\
(Retired from Department of Mechanics, KTH- Royal Institute of Technology) }
\begin{abstract}
Two connected equiperiodic one-dimensional multi-well potentials of different well depths are studied.  Floquet/Bloch energy bands for respective multi-well potential are found to be relevant for understanding level structures. Althoug energies are classically allowed in both multi-well potentials, a band gap of one multi-well potential makes this potential quantum-mechanically 'forbidden'. All energy levels are located in the union of the band regions.
\end{abstract}
\maketitle

\section{Introduction}
Floquet/Bloch-type problems in one dimension can be studied by the amplitude-phase method \cite{T19b}. The method allows calculations of bound states by a Bohr-Sommerfeld-type quantization condition even for multi-well potentials \cite{T15}. 
Single truncated periodic potentials surrounded by vanishing potentials on both sides has also been analysed for scattering energies \cite{T20a}. Floquet characteristics such as energy bands and energy gaps are found to be valuable analytic concepts for explaining energy bands of total transmission \cite{T20b}.

Bound-state problems for double-well potentials  are treated with various amplitude-phase method without the use of Floquet theory \cite{T15}. The doubling of energy levels of such potentials is well known, with larger level splittings for more massive barriers between the wells.  
Generalization to multi-well potentials is simplified if the wells are identical ('truncated periodic potentials'), which allows identification of band/gap energy regions. A further generalization considered, is the combination of two such multi-well potentials attached to each other. Both multi-well potentials are assumed having the same periodicity.

Recent research on multi-well or multi-barrier potentials 
\cite{Dharani16,Yu17,Griffiths01,Sprung93} 
often relates to physical and technical properties of linear nano structures. It
focuses on transmission behaviors of external (charged) particles.  The formalism for multi-well potentials is similar to that for multi-barrier potentials \cite{Yu90,Maiz15,Achilleos17,Shao,Nanda06,Mukhopadhyay12,Karavaev93}. Total transmission was discussed in energy transmission bands known to contain at most $N-1$ peaks of total transmission for $N$-barrier/well potentials \cite{Davydov89}. This assertion is modified in \cite{T20a}, where the number of peaks in a transmission band is shown to be either $N$ or $N-1$, depending on the type of bands considered. Band types was not discussed in earlier publications.

The present study applies the knowledge of (Floquet/Bloch) energy band edges \cite{T19b}, and combines it with a Bohr-Sommerfeld-type wave phase condition for bound states. This condition originate from formulating independent wave functions expressed as $A(x)\cos p(x)$ and  $A(x)\sin p(x)$, where an amplitude function $A(x)$ and a phase function $p(x)$ are defined along an $x$-axis in space. 
The amplitude functions satisfy a Milne-Penny equation \cite{Milne,Wheeler,Pinney} and phases are defined by these amplitude functions. 

Section \ref{sec2} presents the second-order Schr\"{o}dinger equation in non-dimentional form. In section \ref{General} the general Bohr-Sommerfeld condition is derived and energy levels are calculated. Section \ref{Bands} discusses the relevant Floquet/Bloch bands involved.    Conclusions are in section \ref{Conclusion}.

\section{Bound-states} \label{sec2}
The time-independent Schr\"{o}dinger equation with a dimensionless space coordinate $x$ is given by \cite{T19b}
\begin{equation}
F'' + 2m\left(E-V(x) \right) F =0,
\label{ode2} 
\end{equation}
where a prime ($'$) means differentiation with respect to $x$. The dimensionless symbol  $m$ represents an effective mass and equation (\ref{ode2}) is expressed as if being in atomic units. $V(x)$ represents a potential energy function that {vanishes} outside an interval $0\leq x\leq N\pi$, where $N$ is the number of potential periods.   $E (< 0)$ represents the total energy.  Equation (\ref{ode2}), with $V(x)$ being a periodic function of $x$, is a special case of a so-called 'Hill equation' \cite{Hill,McL:eps} . 

Two truncated periodic potentials are connected with $\pi$ being the unit length of a period cell.  The total number of such cells is $N$, an even number, with $N/2$ cells in each multi-well potential. The two locally periodic potentials are fitted to vanishing exterior potentials.  The non-zero Schr\"{o}dinger potential is
\begin{equation}
V(x)= v_1 \sin^2(x),\;\; 0\leq x\leq N\pi/2, V(x)= v_2 \sin^2(x),\;\; N\pi/2\leq x\leq N\pi, \label{poti}
\end{equation}
where $v_1$ is varied and $v_2$ is fixed here. The potential vanishes outside the $x$-region in (\ref{poti}).

Bound-state solutions $F(x)$ satisfy
\begin{equation}
F(x)\to 0, \;\; |x|\to \infty. \label{Lcond}
\end{equation}

An approach using the separate intervals $-\infty<x\leq N\pi/2$ and $N\pi/2 \leq x < \infty$ is presented. Solutions satisfying (\ref{Lcond}) are integrated from $x=N\pi/2$ in two directions using amplitude-phase equations. 
\begin{figure}
\begin{center}
\includegraphics[width=9cm, angle=0]{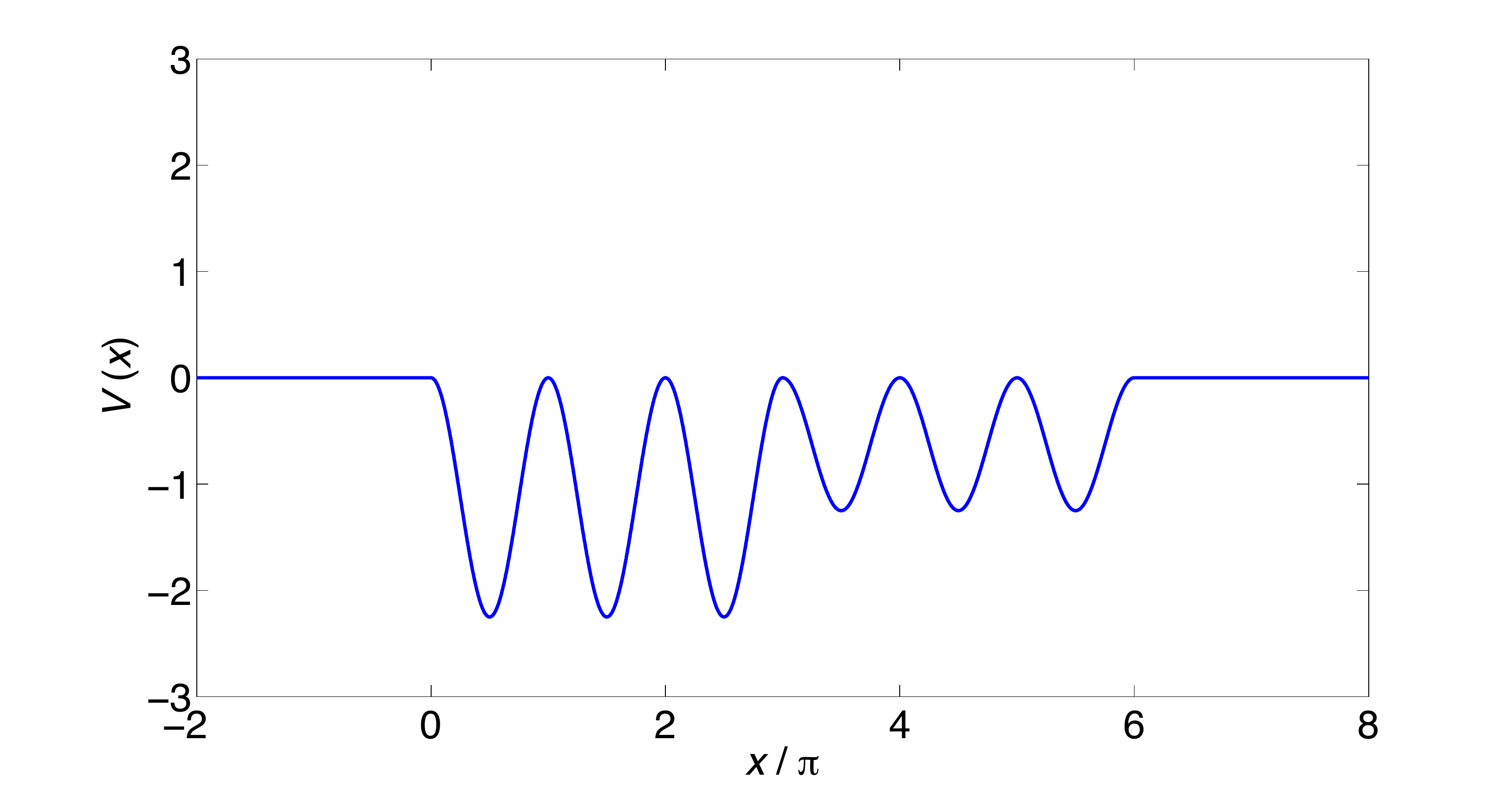}
\caption{Illustration of multi-well potential type considered. The total number of potential cells in the figure is $N=6$. Potential parameters are $v_1=-2.25$ and $v_2=-1.25$.}
\label{fig1}
\end{center}
\end{figure}

\section{General Bohr-Sommerfeld quantization condition for bound states} \label{General}
Amplitude functions satisfy a non-linear differential equation, the Milne-Pinney equation \cite{Milne}-\cite{Pinney}. Any relevant wave function is defined by them.

Two independent wave solutions of (\ref{ode2}) are defined in terms of a positive amplitude function $A(x)$ and a related real phase function $p(x)$ as \cite{T19b} 
\begin{eqnarray}
\Psi^{(\pm)}(x)=A(x) \exp (\pm\rmi p(x)),\label{ansatz}\\ p'(x) = A^{-2}(x) \;(>0). \label{Mephase}
\end{eqnarray}
Relation (\ref{Mephase}) makes sure that the Wronskian of the two solutions (\ref{ansatz}) is independent of $x$, see reference \cite{T05a}. 
An amplitude function satisfies a nonlinear Milne-Pinney equation 
\begin{equation}
A''(x)+2\left[E-V(x) \right]  A(x)={A}^{-3}(x).
\label{Me} 
\end{equation}
Amplitude functions differ by their boundary conditions \cite{T18b} and are more or less oscillating. For molecular masses the choice of slowly varying amplitude functions are typically inspired by the WKB amplitude functions, which are slowly varying in each classically allowed region.

Equation (\ref{Me}) is integrated as a first-order differential equation
\begin{equation}
\left[\begin{array}{c}   A(x)  \\     A'(x) \\   p(x)  \end{array} \right]' 
= \left[ \begin{array}{c}    A'(x)  \\  {A}^{-3}(x)- 2(E-V(x)) A(x) \\  A^{-2}(x) \end{array} \right]. \label{num}
\end{equation}
An integration starts at a boundary point with boundary conditions for the amplitude function. The phase function needs  an additional integration constant to be specified. These equations can be applied in various ways. Often the equations are applied several times. For example in each characteristic region of the potential; i.e. in locally periodic regions and in exterior, asymptotic regions.

The most simple bound-state representation of the Schr\"{o}dinger wave is
\begin{equation}
F(x) = A(x)\sin \phi(x), \label{APgeneral}
\end{equation}
where the phase function $\phi(x)$ includes a specific integration constant, while $p(x)$ in (\ref{num}) has an unspecified integration constant.
For bound states the representation (\ref{APgeneral}) tends to zero as $|x|\to \infty$. Since $A(x)\neq 0$ one requires
$\sin \phi(x) \to 0$ as $x\to -\infty$, and also $\sin \phi(x) \to 0$ as $x\to \infty$. In fact, $A(x)\to \infty$ as $|x|\to \infty$. However, it can be shown that $\sin \phi(x) \to 0$ faster than $A(x)$ diverges in these $x$-limits. Boundary conditions for the quantities in (\ref{num}) are chosen as
\begin{equation}
A(N\pi/2) = 1, \;\; A'(N\pi/2) =0,\;\;  p(N\pi/2) = 0. \label{bcond1}
\end{equation}
The position $x=N\pi/2$ is  between the two locally periodic potentials. A single amplitude function is used. Two integrations start at $x=N\pi/2$ and ends at some sufficiently 'large' negative $x=-x_L$ and at some sufficiently large positive $x=x_R$.
Two phase values are obtained after the integrations:  $\alpha = -p(-x_L)$ and $\beta = p(x_R)$. The relevant phase function $\phi(x)$ in (\ref{APgeneral}) is defined as $\phi(x) = p(x) + \alpha$, and its phase value as $x\to +\infty$ is $\phi(+\infty)=\alpha + \beta$. The bound state condition is
\begin{equation}
\alpha+\beta=(j+1)\pi, \;\; j=0, 1, 2, \cdots. \label{QC}
\end{equation}
In this way the wave function $F(x)$ is normalized to satisfy $F(N\pi/2)=\sin \alpha$ and $F'(N\pi/2)=\cos \alpha$.
\begin{figure}
\begin{center}
\includegraphics[width=12cm, angle=0]{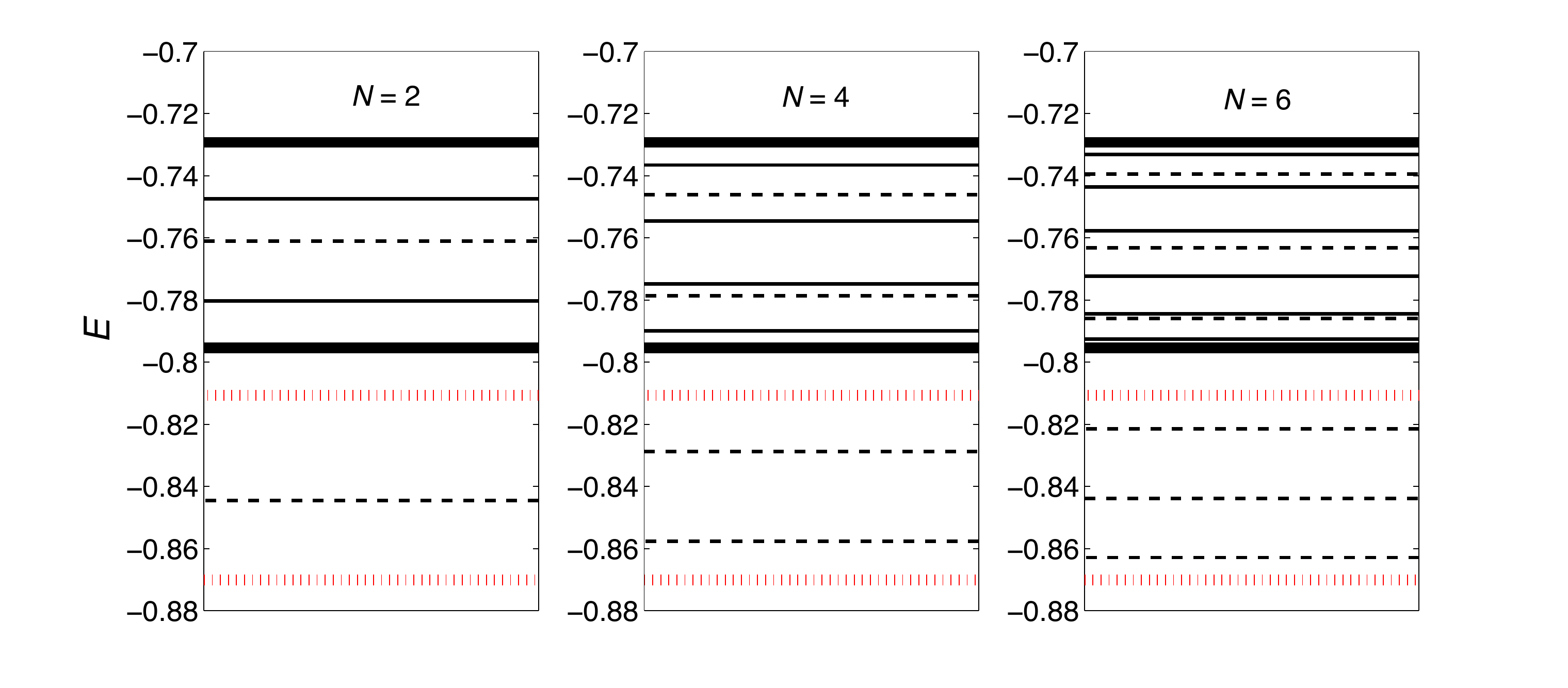}
\caption{Levels corresponding to the two potential cases $(v_1,v_2)=(-1.25,1.25)$ (solid lines) and $(v_1,v_2)=(-1.35,1.25)$ (dashed lines). The broader solid (black) and dashed (red) lines indicate edges of the two Floquet/Bloch bands involved.}
\label{fig2}
\end{center}
\end{figure}
\begin{table}
\begin{center}
\small
\begin{tabular}[t]{cccccc}
\hline
\hline
$j$&$(v_1,v_2)=(-1.25,-1.25)$&$E_j$&$(v_1,v_2)=(-1.35,-1.25)$&$E_j$& \\
\hline
&$N=2$&&&\\
\hline
&Band edge  $E=-$0.7293&&\\
\hline
1&&-0.7475&&-0.7610 \\
0&&-0.7803&& \\
\hline
&Band edge  $E=-$0.7953&&\\
\hline
\hline
 &Band edge  $E=-$0.8106&&\\
\hline
1&&&&\\
0&&&&-0.8445\\
\hline
&Band edge $E=-$0.8701&&\\
\hline
\hline
\end{tabular}
\caption{\small Bound state energy levels for $N=2$ corresponding to Figure 1. The case $(v_1,v_2)=(-1.25,-1.25)$ corresponds to a symmetric double-well potential. The case $(v_1,v_2)=(-1.35,-1.25)$ corresponds to an asymmetric double-well potential.}
\end{center}
\label{table1}
\end{table}
\begin{table}
\begin{center}
\small
\begin{tabular}[t]{cccccc}
\hline
\hline
$j$&$(v_1,v_2)=(-1.25,-1.25)$&$E_j$&$(v_1,v_2)=(-1.35,-1.25)$&$E_j$& \\
\hline
&$N=4$&&&\\
\hline
&Band edge  $E=-$0.7293&&\\
\hline
3&&-0.7366&&-0.7461 \\
2&&-0.7546&&-0.7787 \\
1&&-0.7748&& \\
0&&-0.7898&& \\
\hline
&Band edge  $E=-$0.7953&&\\
\hline
\hline
 &Band edge  $E=-$0.8106&&\\
\hline
1&&&&-0.8107\\
0&&&&-0.8577\\
\hline
&Band edge $E=-$0.8701&&\\
\hline
\hline
\end{tabular}
\caption{\small Bound state energy levels for $N=4$ corresponding to Figure 1. The case $(v_1,v_2)=(-1.25,-1.25)$ corresponds to a symmetric quadruple-well potential. The case $(v_1,v_2)=(-1.35,-1.25)$ corresponds to an asymmetric quadruple-well potential.}
\end{center}
\label{table2}
\end{table}

The method resulting in (\ref{QC}) is exact. It is numerically best suitable for light-particle interaction potentials with effective masses $m$ satisfying $(0)<m<5$, and cell numbers satisfying $(0)\leq N<<10$. The time efficiency is gradually lost for larger values of $m$ and $N$.

Figure 2 shows energy levels for two multi-well combinations, $(v_1,v_2)=(-1.25,1.25)$ (solid lines) and $(v_1,v_2)=(-1.35,1.25)$ (dashed lines), and $N/2$ cells for each well depth ($v_1$ respectively $v_2$). 
 Hence, the solid level lines correspond to single-type $N$-well potentials; $N=2$ corresponding to an ordinary double-well potential, $N=4$ to an ordinary quadruple-well potential, and so forth. As $N$ increases, the levels become more and more dense. However, the levels are contained in a finite range of energies, a particular Floquet/Bloch band region corresponding to $v_1$ or $v_2=-1.25$. 

The dashed levels corresponding to $(v_1,v_2)=(-1.35,1.25)$, two different multi-well potentials, also become more dense as $N$ increases. They appear in two groups, contained in the two relevant Floquet/Bloch bands, one band corresponding to $v_1$, and one band corresponding to $v_2$. These energy bands do not overlap. Energy levels are not observed between the band regions.

\section{Floquet/Bloch bands} \label{Bands}
Any periodic potential with a given value of the strength  has specific energy Floquet/Bloch bands. 
Each value of $v_1$ and/or $v_2$ is associated with several, more or less separated, energy bands. 
For single-well potentials corresponding to either $v_1$ or $v_2$, one has two sets of bound-state levels. Each level in each potential develops into a Floquet/Bloch band, as more and more identical wells are added. The phenomenon is a kind of discrete multi-well level splitting, approaching a continuous energy band as the number of identical wells tends to infinity.

The resulting bands corresponding to $v_1$ and $v_2$ may, or may not, overlap.
Only the bands corresponding to ground state levels of the single wells are considered here. 

Energy band edges satisfy any of the conditions \cite{T19b}
\begin{equation}
u\cos \gamma = \pm 1. \label{BSC}
\end{equation}
The symbol $u$ represents an amplitude value after integrating (\ref{num}) accross one cell (a single well) of the potential. Boundary conditions for the amplitude function are $A(x_b)=1$, $A'(x_b)=0$ and $p(x_b)=0$. The position '$x_b$' is chosen as either '$0$', for the left multi-well potential in Figure 2,  or '$N\pi/2$', for the right multi-well potential in Figure 2. The phase $\gamma$ represents the value of $p(x)$ in (\ref{num}) obtained ofter integration at $x=x_b+\pi$.
The band edges satisfy $u\cos \gamma = 1$ or $u\cos \gamma = -1$.

The main observation is that, althugh the number of cells is small, the band concept is still relevant. No energy levels of the present multi-well potentials are located in {\it between} the relevant bands.
\begin{table}
\begin{center}
\small
\begin{tabular}[t]{cccccc}
\hline
\hline
$j$&$(v_1,v_2)=(-1.25,-1.25)$&$E_j$&$(v_1,v_2)=(-1.35,-1.25)$&$E_j$& \\
\hline
&$N=6$&&&\\
\hline
&Band edge  $E=-$0.7293&&\\
\hline
5&&-0.7331&&-0.7396 \\
4&&-0.7436&&-0.7632 \\
3&& -0.7577&&-0.7859 \\
2&&-0.7723&& \\
1&&-0.7845&& \\
0&&-0.7925&& \\
\hline
&Band edge  $E=-$0.7953&&\\
\hline
\hline
 &Band edge  $E=-$0.8106&&\\
\hline
2&&&&-0.8215 \\
1&&&&-0.8438 \\
0&&&&-0.8628 \\
\hline
&Band edge $E=-$0.8701&&\\
\hline
\hline
\end{tabular}
\caption{\small Bound state energy levels for $N=4$ corresponding to Figure 1. The case $(v_1,v_2)=(-1.25,-1.25)$ corresponds to a symmetric quadruple-well potential. The case $(v_1,v_2)=(-1.35,-1.25)$ corresponds to an asymmetric quadruple-well potential.}
\end{center}
\label{table3}
\end{table}
\section{Concluding remarks} \label{Conclusion}
The multi-well potentials in this study are locally periodic and can be associate with band/gap properties.
One finds that a multi-well potential attached to another multi-well potential sees its neigbor potential as a 'forbidden' region at certain energies, even if the energy is classically allowed in both potentials.   The energy then happens to lie in a band gap of the neighbor potential. A consequence is a small shift of energy levels compared to the unconnected multi-level potential. As the multi-well potentials become identical, all levels are confined to a single Floquet/Bloch energy band.

A feature of the amplitude-phase approach for quantal waves is that amplitude functions can be defined by different boundary conditions without changing the original quantal wave. Other approaches may explore this 'amplitude freedom' to treat characteristic regions (exterior or locally periodic ones) by completely separate amplitude-phase representations. Such alternative formulations lead to more complicated quantization conditions. On the other hand such conditions become more time efficient than the present approach for large values of $m$ and $N$.


\end{document}